# Indirect dark matter searches in Gamma- and Cosmic Rays

Jan Conrad[1] and Olaf Reimer[2]

1 Oskar Klein Center, Physics Department, Stockholm University, Roslagstullsbacken 21, Stockholm, Sweden.

2 Institute for Astro- and Particle Physics, Innsbruck University, Technikerstr 25/8, Innsbruck, 6020, Austria.

Dark matter candidates such as weakly-interacting massive particles are predicted to annihilate or decay into Standard Model particles leaving behind distinctive signatures in gamma rays, neutrinos, positrons, antiprotons, or even anti-nuclei. Indirect dark matter searches, and in particular those based on gamma-ray observations and cosmic ray measurements could detect such signatures. Here we review the strengths and limitations of this approach and look into the future of indirect dark matter searches.

______________________________________________________________________

Evidence for particle dark matter is so far circumstantial. Cosmological observations on different scales (see e.g. Bergström 2012) are most convincingly explained by introducing new, as yet unknown, particles whose existence at best also solves other conundrums in modern physics. Proposed particle candidates for dark matter span more than 60 orders of magnitude in cross-section (with Standard Model particles) and about 45 orders of magnitude in mass (Fig.1). Obviously, no single experimental technique can cover such a parameter range. The most popular candidate, the very focus of this article, is a particle type that is weakly interacting, but much more massive than a neutrino (Weakly Interacting Massive Particle, or WIMP). A very plausible hypothesis for the production of dark matter is that it consists of thermal relics of the Big Bang (much like the photons of the cosmic microwave background). Since the annihilation of dark matter particles into Standard Model particles controls the abundances in the Universe, principally one expects a firm connection between their ability to interact—expressed by the velocity-averaged annihilation cross section $\langle \sigma v \rangle$, which for briefness we will simply refer to as annihilation cross-section)—and the cosmologically relevant properties or observables. In particular, the abundance of thermal relics can be accurately calculated and yield, with two simple assumptions on interaction strength and mass for a WIMP, an abundance that is close to the one that is very accurately measured by cosmological observations. Sometimes this coincidence is popularized as the 'WIMP miracle'. As the abundance is regulated by the already mentioned annihilation cross-section, requiring that the relics provide the entire observed dark matter provides a benchmark for indirect detection at about $3 \times 10^{-26}$ cm$^3$ s$^{-1}$.

An additional feature of WIMPs is that particle theories beyond the Standard Model, invoked for a different reason than dark matter, often generically include a WIMP. In particular, Supersymmetry, which roughly doubles the number of particles in the Standard Model, provides such a particle, referred to as neutralino. Self-annihilation is consequently a viable scenario to be investigated by indirect dark matter search techniques.

http://www.nature.com/nphys/journal/v13/n3/fig_tab/nphys4049_F1.html

**Figure 1:** Dark matter candidates indicating the interdependence of the interaction cross-section and particle mass. The candidate most generically within reach of indirect detection belongs to the concept of Weakly Interacting Massive Particles (WIMPs), predicted by a variety of theories, most notably Supersymmetry, i.e. the neutralino. KK stands for Kaluza-Klein, LTP refers to lightest T(ime-parity)-odd particle and CDM is cold dark matter. Figure reproduced from Park 2007.

Apart from establishing the particle nature of dark matter an unambiguous indirect detection of dark matter annihilation will also yield information about its microscopic properties. In particular, detection would allow for the estimation of the annihilation cross-section (assuming known dark matter velocity and density distribution) and the dark matter mass. In the absence of a clear signal, constraints on these quantities can still be obtained. To a lesser extent, but still of importance, the signature in gamma- and cosmic-ray observations is determined by the composition of Standard Model particles that the WIMP annihilate. Most often single annihilation channels are assumed (for instance annihilation into b-quark pairs), mainly motivated by the fact that single channel annihilation can serve as a representative for a larger class of models. Some works instead compute the gamma-ray yield for the full set of annihilation channels and branching fraction as predicted from the underlying theory, e.g. supersymmetry, pioneered in Scott et al. 2010.

Indirect dark matter searches rarely constitute the one and only scientific objective of experiments designed to observe cosmic rays or photons at the upper end of the electromagnetic spectrum. More commonly, putative dark matter signatures are investigated while performing rather conventional cosmic ray intensity or composition measurements or when surveying the skies in gamma-rays. Indirect dark matter searches are carried out whenever opportunities promise to be sufficiently sensitive in the respective search windows and anticipated parameter space of dark matter candidates. Since dark matter is generally thought to be of universal nature, the **universality** in conducting indirect dark matter searches connects inevitably with consistency to support the credibility of the findings. Reported anomalies (that is potential or weak indications of a dark matter signature) can be relatively easily strengthened or refuted when further studied in different search regions or using alternative search methods, without the need to build up alternative experiments to do so. The requirements for **consistency** between different applications of indirect dark matter searches connect seamlessly to aspects of **complementarity** with other approaches to the detection of dark matter particles, such as direct detection (observation of the scattering of dark matter particles in underground experiments) and/or collider searches for dark matter.

The discovery through indirect detection techniques requires a profound understanding of both the astrophysical backgrounds and the dark matter model-imposed uncertainties. In addition, predictions for the reach of indirect detection are plagued mainly by the uncertainty in the dark matter spatial distribution. The expected flux from dark matter annihilations is proportional to the integral of the dark matter density squared over the line of sight and the solid angle subtended by the observation. This integral is traditionally referred to as the 'J-factor'. The uncertainties in the density distribution consequently lead to systematic uncertainties that range between factors of a few to orders of magnitude dependent on what target is chosen. For decaying dark matter, the respective cross-section enters linearly, with the corresponding integral being sometimes referred to as 'D-factor'.

These uncertainties *per se* do not impact on the credibility of any discovery. However, an additional challenge for indirect detection is the fact that astrophysical sources, especially in the usual regime of limited statistics can mimic sources of dark matter annihilation Whereas direct detection also suffers from (comparably smaller) astrophysical uncertainties mainly in the dark matter velocity distribution and local dark matter density, direct detection appears to be the most straightforward method for discovery, leaving the credibility only subject to the ambiguity in controlling the experimental setups and instrumental backgrounds. Particle collider searches can discover dark matter candidates and once the connection is made between these candidates and cosmological dark matter, they have the chance to elucidate the properties of dark matter. However, once again, owing to the uncertain nature of the potential interaction channels, collider searches might still fail even if the mass range would suffice. Finally, indirect dark matter search techniques can benefit from **serendipity**. Discoveries in the high-energy universe have the potential to reveal places with extremely promising characteristics for dark matter studies, and the indication of anomalies, interpreted as potential dark matter signatures, may arise as the by-product of studying other astrophysical phenomena. The history of discoveries in astronomy, cosmology and astroparticle physics testifies that serendipity, while unable to deploy into an active search method, did bring substantial insights.

## How to search for dark matter using indirect methods?

There are a variety of anticipated experimental signatures of particle dark matter that leave imprints in the observable energy spectra and/or spatial distribution of gamma-ray photons or charged cosmic rays. Statistical techniques to exploit such signatures —foremost the multi-dimensional profile likelihood and template fitting signal decomposition—have had significant impact on the progress of indirect detection.

A principal challenge for indirect detection methods is the issue of source confusion and poorly determined backgrounds. It is well known that both for the gamma-ray and the charged cosmic ray channel pulsars provide spectral signatures that are in most practical cases indistinguishable from dark matter. So far the only known smoking-gun signal indirect detection can provide is therefore based on the unique spectral features, the most spectacular being a spectral line originating from the annihilation of dark matter particles with each other resulting in either two photons or a boson and a photon (or both, for multiple lines) (Srednicki et al. 1986, Bergström et al. 1988). Generically, such processes are suppressed as they are almost exclusively possible via loop processes, but different mechanisms can lead to enhancements (Hisano et al. 2004). Distinctive spectral features not only provide a smoking-gun signal, but they also allow the experimentalist to choose a data-driven method for inferring the background, as control regions are easily defined in this case.

There are celestial regions where dark matter searches appear more promising. As detailed below, this relates to the anticipation of the successful distinction between dark matter-related emission signatures and the omnipresent astrophysical backgrounds. When exploring over-densities in gravitationally-bound matter, the regular morphology of dark matter-related signals turns out to be a powerful discriminator against usually unevenly-structured astrophysical emissions. N-body simulations of the cosmic structure allow for the prediction of spatial mass density profiles, with the common features among them being smooth and regular density gradients away from a central mass or mass assembly, parametrized as the NFW, Einasto, Moore, or Burkert-DM-halo density distributions (Navarro et al. 1996, Einastro 1965, Moore et al. 1999, Burkert 1995).

An indirect method seeking for evidence for dark matter annihilation on cosmological scales (Ando & Komatsu 2006) measures the cross-correlation between astronomical object catalogs (Xia et al. 2011, Ando et al. 2014, Cuocu et al. 2015) or gravitational distortion in the weak lensing regime (Camera et al. 2013, Shirasaki et al. 2014) and the extragalactic gamma-ray background. Whereas a positive correlation is the principal evidence for the cosmological origin of the extragalactic gamma-ray background, cross-correlation signals originating from dark matter annihilation are anticipated to be different from those of astrophysical foregrounds. The intensity, spectrum, and spatial distribution of resolved and unresolved gamma-ray sources as well as large-scale galactic emission (Hopper et al. 2007, Cuoco et al. 2008) leave imprints on different angular scales than those of annihilating dark matter particles. The degeneracy between different scenarios and contributions is anticipated to be reduced when the angular cross-correlation is investigated by considering a multitude of astronomical object catalogs, and in different energy windows. Another way to investigate the extragalactic gamma-ray background for dark matter-induced angular features (anisotropies) on the cosmological scale emerged by considering the auto-correlation angular power spectrum (Ackermann et al. [Fermi-LAT] 2012, Ando & Komatsu 2013, Fornasa et al. 2013). The predicted shape of the angular power spectrum of gamma-rays originating from dark matter annihilation deviates from that caused by other astrophysical sources where intensity and density scale linearly. Guaranteed contributions from unresolved sources to the extragalactic gamma-ray background, as well as astrophysical foregrounds leaving imprints in the angular power spectrum render the interpretation of the results from this method strongly conditional on the assumptions of the analysis methodology.

Experimental techniques in cosmic ray physics offer sufficiently precise measurements of the charge, charge-sign, momentum and mass to identify individual cosmic ray particles or nuclei over a large energy range. This energy scale conveniently corresponds to the mass range of WIMPs. Anomalies in cosmic ray spectra, or more precisely in the measurements of antiparticles like antiprotons or positrons, as well as sensitive limits on heavier antinuclei, are explored for contributions potentially originating from the annihilation and decay of dark matter particles into pairs of Standard Model particles, subsequently decaying or hadronizing into particles that blend with the cosmic rays from astrophysical sources. The major background in these measurements is no longer the misidentification probability to cosmic ray particles and heavier nuclei or event statistics, but the distance-, time- and energy dependence of cosmic ray sources and the propagation leaving an imprint on their the relative intensities in a complex way. The excess flux against that predicted from standard scenarios for the origin and transport of Galactic cosmic rays could either be interpreted as the imprint from one (or more) sources that supply electrons and positrons to the interstellar medium (Aharonian et al 1995, Nishimura et al. 1997, Kobayashi et al 2004, Aharonian & Atoyan 1991), or from dark matter annihilation in the TeV range. Inadequacies in modeling of cosmic ray transport seem to prevent solving this dichotomy presently. Also the antiproton spectrum is studied for deviations from the pure secondary production in cosmic ray interactions. In the light of the recent Alpha Magnetic Spectrometer (AMS) data, the situation is even more ambiguous, as recent refinements of the primary cosmic ray spectra and cross-sections for the calculation of secondary particle production already ease the apparent tension with conventional scenarios (Cirelli 2016). Other antinuclei, e.g. antideuterium, antihelium etc., have not ever been detected. These observations might, however, become very powerful probes for dark matter searches as the ambiguity regarding the astrophysical backgrounds is mostly absent. Still, the experimental limits are orders above even the most optimistic predictions.

Charged cosmic rays at sub-TeV energies are assumed to be isotropized in their arrival direction when they reach the Earth, with the potential exception of electrons and positrons which could be indicating the presence of a nearby source. Anisotropies in the cosmic ray flux measured on Earth can be investigated for consistency with proposed dark matter scenarios, e.g. the observed arrival directions of high-energy electrons and positrons can be compared to those from alternative astrophysical source scenarios. When comparing the cosmic ray anisotropy signatures (Profumo 2015) or the rising positron fraction with gamma-ray observations (Lopez 2016), strong constraints on the dark matter-related interpretations can be obtained. The interpretation of the intriguing TeV-scale hadronic cosmic ray anisotropy (Amenomori et al. [Tibet AS$\gamma$], Abdo et al. [MILAGRO], Aartsen et al. [IceCube], Abeysekara et al. [HAWC]) in terms of annihilating dark matter is considered to be problematic (see Ahlers & Mertsch 2017 for a recent review).

To dissect the cosmic ray measurements regarding their relation to either conventional or dark matter-induced astrophysical processes we require a better understanding of the cosmic ray transport in our Galaxy, either by accessing more realistic propagation scenarios, invoking improved models for radiation fields or refined matter distributions in our Galaxy, and more complete as well as more precisely-measured cross-sections for kinematic interactions of cosmic rays.

The most frequently applied indirect dark matter search technique relies on a given set of high-level observational data (e.g. gamma-ray skymaps) which are then re-analyzed by adding dark matter-specific spatial distribution templates. The Large Area Telescope aboard the Fermi Gamma-ray Space Telescope (Atwood et al. [Fermi-LAT] 2009) is presently the prime instrument delivering input (Acero et al. [Fermi-LAT] 2015) for signal decomposition techniques, thanks to its large field-of-view, multi-year exposure, and broad dynamic regime in the gamma-rays. Likewise, residual emission features from a given gamma-ray analysis might be further studied, e.g. by comparison with model-predicted dark matter annihilation or decay signatures. Improvements in the template decomposition techniques are often accomplished through iterative procedures where a suitable statistical estimator is used to quantify the improvements in the results. The most commonly practiced approach involves a pixel-wise Poisson likelihood (see also next section).

Apart from the gain in instrumental sensitivity which is usually accomplished by increasing the exposure or improvements in the event reconstruction—i.e. the mapping between the electrical signals in the detector and the physical properties, as well as the classification of the events into certain particle types or interaction categories—the application of dedicated statistical techniques has led to significant improvements in sensitivity. In particular, Fermi-LAT has implemented a multi-dimensional likelihood analysis which paved the way for the optimal target combination and statistically more accurate treatment of nuisance parameters, e.g. the dark matter density estimate by means of the profile likelihood. In this frequentist technique, the observables' dependence on ancillary (nuisance) parameters is modelled and the parameter estimates are obtained by maximizing the likelihood with respect to them. The profile likelihood has been known in the high energy physics community for at least thirty years but gained popularity with the advent of the Large Hadron Collider

(LHC) in the last decade (James 1980, Rolke et al. 2005, Cowan et al. 2011). Its application to Fermi-LAT observations of dwarf galaxies lead to the first exclusion of thermally-produced WIMPs (for masses below 30 GeV) as being the dominant part of dark matter (Ackermann et al. [Fermi-LAT] 2011). The multi-dimensional likelihood approach has then also found its way to searches performed by Imaging Atmospheric Cherenkov Telescopes. The recent very competitive constraints obtained by the HESS collaboration (Abdallah et al. [HESS] 2016) exemplify the power of this approach. It is worth mentioning that similar techniques are also applied in direct searches for of WIMPs (see Conrad 2015 for a recent review).

http://www.nature.com/nphys/journal/v13/n3/fig_tab/nphys4049_F2.html

**Figure 2:** Targets for indirect dark matter searches in the gamma-ray sky. The central Fermi-LAT skymap indicates the celestial distribution of high-energy photons. Symbolizing one or more specific characteristics of a respective search location, the most popular targets are emphasized in auxiliary pictures and discussed in the text. By GC we denote the Galactic Centre and by dSph dwarf spheroidal galaxy.

## Where to search for dark with indirect methods?

The detectability of dark matter signatures in gamma-rays is proportional to the J-factor introduced before. Obvious locations for indirect dark matter searches are therefore regions of extraordinary matter density (e.g. centres of gravitationally bound bodies or assemblies such as galaxies or galaxy clusters), peculiar objects or regions with high dark matter content (clumps, sub-halos, dwarf spheroidal galaxies), and the most prominent large-scale diffuse emission phenomena in the gamma-ray sky. For consistency, indirect dark matter searches explore a range of different targets (Fig.2).

The largest assembly of matter in our galaxy is naturally the most searched for location for dark matter signatures. However, it is also the most challenging location to investigate given the variety of astrophysical objects in the inner Galactic Centre region, ranging from a unique object like our central black hole Sag A*, to a number of Supernova Remnants, neutron stars and pulsars. Furthermore, all diffuse emission components in the gamma-rays (neutral pion decay, inverse Compton scattering, bremsstrahlung) peak in the Galactic Centre. The brightest and central gamma-ray point-source in the region is extensively searched for line-like spectral features and limits have been placed over both GeV and TeV energies (Abdo et al. [Fermi-LAT] 2010, Ackermann et al. [Fermi-LAT] 2012, Ackermann et al. [Fermi-LAT] 2015, Abramowski et al. [HESS] 2013, Abramowski et al. [HESS] 2015, Abdallah et al. [HESS] 2016). It is now believed that regularly extended emission centered at the Galactic Centre holds more promise for indirect dark matter searches. Indeed, the so-called

'Galactic Centre excess' (Goodenough & Hooper 2009, Hooper & Goodenough 2011, Hopper & Slayter 2013, Daylan et al. 2016, Abazajian et al. 2012, Calore et al. 2015, Gordon & Macias 2016) is still not conclusively explained, although recently improved astrophysical source models, including the unresolved stellar population render hypothesis for a unique dark matter explanation of the observed GeV Galactic Centre excess signal, are less likely (Calore et al. 2015, Ajello et al. [Fermi-LAT] 2016, Yang & Aharonian 2016, Macias 2017). Millisecond pulsars might be able to account for this anomaly. Since they exhibit a broadband emission spectrum, deeper surveys, e.g. at radio wavelength, might enlarge this object's distribution and shed further light on this anomaly. However, there are contrasting views on whether the population of millisecond pulsars is sufficiently large to account for the observed gamma-ray signal (Cholis et al. 2015).

Searching the extended celestial regions is most likely not limited anymore by the low number of associated photons, but by the systematic uncertainties of the deployed analysis procedure or intrinsic to the experimental detection technique of photons itself. The spatially symmetric and extended gamma-ray halo around the centre of our Milky Way is one of the foremost locations for dark matter signatures. Such a structure is a common feature among the different models for spatial mass density profiles predicted by N-body simulations of the cosmic structure. In the Milky Way, establishing the potential existence of an extended halo is challenging for reasons of inferring the astrophysical foregrounds at the various scales. Resolved sources, point-like or extended, large scale emission phenomena as predicted for the Inverse Compton component in the Diffuse Galactic Emission, the existence of large bipolar outflows (or 'Fermi-bubbles') and populations of still unresolved sources of known or even unknown nature add to complications in extracting robust measurements of a dark matter-related Milky Way Halo component. The most prominent galactic emission feature, the Diffuse Galactic Emission (e.g. Stecker 1970, Bertsch et al. [EGRET] 1993, Strong et al. 2000), has been repeatedly searched for anomalies (e.g. Ackermann et al. [Fermi-LAT] 2012). Manifestations of imperfect instrumental responses or deficits in the realism of emission models constructed to predict the spatial- and spectral intensity of the reported Diffuse Galactic Emission became known. Being a demanding analysis on its own due to the manifold simplifying assumptions and limitations of the models to forecast the Diffuse Galactic Emission in the observables (e.g. spectrum, longitudinal and latitudinal profiles, tangents and arm/inter arm contrast etc.), the potential signatures of dark matter annihilation in the Diffuse Galactic Emission so far failed to stand tests of universality and consistency over time and/or independent measurements or observation channels.

The Milky Way halo is also the principal source of dark matter induced charged cosmic rays, in particular species rarely produced by conventional mechanisms, such as positrons or antiprotons. Being charged these particles are deflected by the intragalactic magnetic field and do not retain directional information accurately enough to allow the identification of individual sources, except if the source happens to be very close-by. Only in the latter case the asymmetry on the scale of the arrival hemispheres can shed light on the validity of such a scenario. In addition, due to the efficient energy loss process, charged cosmic rays also sample a relatively local volume in the Galactic Halo.

N-body simulations predict a large amount of substructure in a Galaxy-size Milky Way halo in the form of relatively localized density enhancements ('clumps'). Dwarf spheroidal galaxies are a manifestation of such clumps which apart from dark matter exhibit light emission (mostly from gravitationally bound stars). It is however conceivable that part of these clumps would be visible mainly by their dark matter annihilation signal and potentially explain a fraction of those gamma-ray sources that presently remain unidentified. The non-detection of clumps can be used to constrain the microscopic properties of dark matter particles, but, the abundance of dark matter clumps has to be modelled. Dwarf spheroidal galaxies being precisely located targets for dark matter searches provide a direct means to search and constrain dark matter particles. From measurements of the dynamics of stars it is known that these objects are dominated by dark matter with mass-to-light ratios of up to a thousand. Still, the total density contributing to a hypothetical dark matter annihilation signal is one to two orders of magnitude smaller than for the galactic center. Nevertheless, dwarf galaxies are by now the most promising target for WIMP searches for two reasons: the absence of any significant intrinsic diffuse emission or source confusion; and the taming of the main systematic plaguing the inference of particle properties, such as the knowledge of the dark matter density, by using spectroscopy of the member stars which allows for the estimation of the dark matter distribution in these objects. As mentioned above, systematic uncertainties in the cross-section for example are arguably a factor 2-4 (Ullio & Valli 2016).

Quite naturally, Galaxy Clusters, as the largest gravitationally bound structures in the Universe, are also targets for indirect dark matter searches. Although principally considered dark-matter rich, uncertainties relating to the size and alignment of large dark matter halos as well as the number and distribution of small sub-halos in a cluster or the cluster vicinity introduce problems in interpreting even sensitive measurements. This leads again to orders of magnitude of systematic uncertainty in the estimates or constraints on the annihilation cross section, which—in absence of uncertainties—would be competitive with the limits from dwarf galaxies. For decaying dark matter, due to the large mass and only linear dependence on density, Galaxy clusters are very competitive targets. Again, guaranteed astrophysical foregrounds from the cosmic ray interactions taking place in galaxy clusters are believed to outshine any such dark matter-induced gamma-ray emission component. Worse, not even the gamma-ray emission from cosmic ray interactions is being manifested today, neither in individual clusters (Ackermann et al. [Fermi-LAT] 2016a, Ackermann et al. [Fermi-LAT] 2016b, Ahnen et al. [MAGIC] 2016) nor in large sample studies (Ackermann et al. [Fermi-LAT] 2014). Consequently, the limits on the dark matter properties from galaxy clusters are conditional on model-dependent uncertainties in the principal emission components. The first detection of cluster-related gamma-ray emission, currently pending at GeV and TeV energies, would profoundly enhance the quality of the interpreted gamma-ray limits from galaxy clusters. When coinciding with the substantial progress regarding the constraints on the amount of substructure, galaxy clusters might become one of the best targets for indirect dark matter searches.

The residual gamma-ray signal obtained after accounting for all resolved sources and sufficiently motivated galactic foreground emission phenomena is called the Extragalactic Gamma-ray Background . It is explained by the cumulative contributions of unresolved sources in source classes that should be able to emit gamma rays at the large galactic scale heights, such as halo-populations like millisecond pulsars, blazar-class active galactic nuclei, misaligned active galactic nuclei, star-forming galaxies, or galaxy clusters. These constituents are considered as guaranteed contributions to the observed Extragalactic Gamma-ray Background signal, albeit being predicted with different level of uncertainty per class and varying prominence over energy compared to each other. Uncertainties in the predictions of the guaranteed astrophysical contributions to the Extragalactic Gamma-ray Background as well as the potential existence of anisotropies offer chances to explore the Extragalactic Gamma-ray Background for dark matter annihilation or decay signatures. The respective probes (energy spectrum, cross- and/or auto-correlation angular power spectrum) have been already introduced. Expectations to reveal unambiguous signatures of dark matter in the Extragalactic Gamma-ray Background signal relies in a subtle way on both the extremely elaborate analysis procedure to robustly measure the Extragalactic Gamma-ray Background itself (Ackermann et al. [Fermi-LAT] 2015, Sreekumar et al. [EGRET] 1998, Keshet et al. 2004, Strong et al. 2004, Abdo et al. [Fermi-LAT] 2010 and a precise understanding of the contributions from unresolved source populations (Ackermann et al. [Fermi-LAT] 2012, Siegal-Gaskins et al. 2011, Ajello et al. [Fermi-LAT] 2015, Di Mauro & Donato 2015). To a certain degree, the measurement of the Extragalactic Gamma-ray Background can then be used to constraint the intensity of potentially dark matter-related emission (Ullio et al. 2002) and allow placing upper limits on characteristic quantities such as the annihilation cross-section. However, these constraints not only depend on the distribution of dark matter in halos, but also on the abundance of halos and sub-halos and their redshift dependence. Nevertheless, present limits, approaching the most stringent existing constraints, already allow for important consistency checks with those obtained using a different analysis methodology (see Fornasa & Sanchez-Conde 2015 for a recent review).

## Where are we now?

Indirect detection has provided a number of intriguing indications of a dark matter signal, which usually subsequently disappeared —mostly due to the aforementioned systematic uncertainties, difficult backgrounds and possible source confusion. At present, there are only three anomalies that can be ascribed to dark matter: the rising fraction of positrons as compared to electrons measured by PAMELA (Adriani et al. [PAMELA] 2009) and confirmed by AMS (Aguilar et al. [AMS] 2014), a hardening of the antiproton fraction reported by AMS (Aguilar et al. [AMS] 2016) and the excess emission of the Galactic Bulk in the GeV gamma rays measured by Fermi-LAT. There is ample literature discussing the influence of systematic uncertainties and possible conventional astrophysical sources explaining these anomalies. Here, it suffices to say that none of these anomalies lack a plausible conventional explanation and these indications are ambiguous at best. Turning to limits, the currently strongest constraints on the annihilation cross-section and WIMP mass (Fig.3) come from the analysis of dwarf galaxies (Ackermann et al. [Fermi-LAT] 2015). This analysis excludes annihilation cross-sections larger than the thermal cross-section benchmark for dark matter candidates lighter than 100 GeV.

Complementary limits (e.g. Giesen et al. 2015, Jin et al. 2015) obtained from recent antiproton measurements (Aguilar et al. [AMS] 2016) substantiate these tight bounds. Likewise, the observations of the central galactic bulge provide constraints on level pegging (Yang & Aharonian 2016). At higher energies, H.E.S.S. observations of the inner Galactic Halo presently impose the best limits (Abdallah et al. [HESS] 2016).

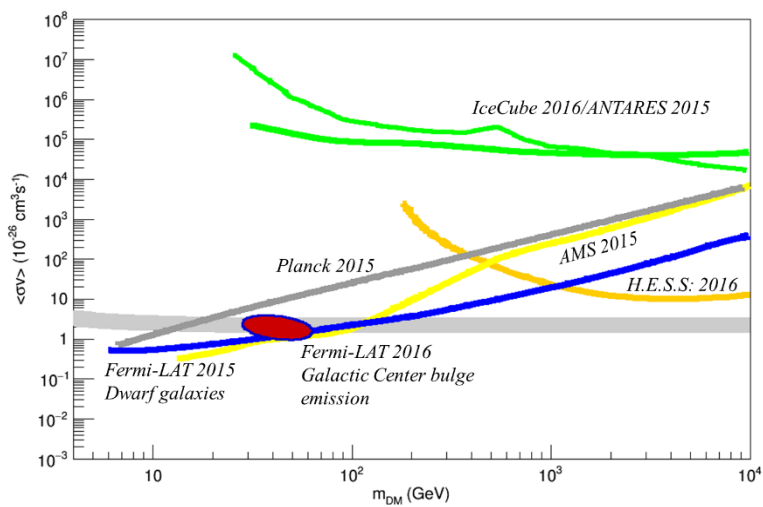

**Figure 3:** The current most important constraints on the annihilation cross-section versus WIMP mass. The constraints are for the annihilation to b-quark pairs. Whereas indirect methods exploring gamma-ray photons and cosmic rays from satellite measurements compete well up to hundreds of GeVs, at higher energies Air Cherenkov Telescopes appear to be driving the present limits. The thermal relic cross-section is indicted by the light grey band. Note that different assumptions for the DM distributions affect these limits quantitatively, but do not change the situation qualitatively.

Data from Ade et al. [PLANCK] 2016, Adrian-Martinez et al. [ANTARES] 2016, Aartsen et al. [IceCube] 2016 531, Ackermann et al. [Fermi-LAT] 2015; Abdallah et al. [HESS] 2016; Gordon & Macias 2013; Daylan et al. 2016, Calore et al. 2015, Giesen et al. 2015

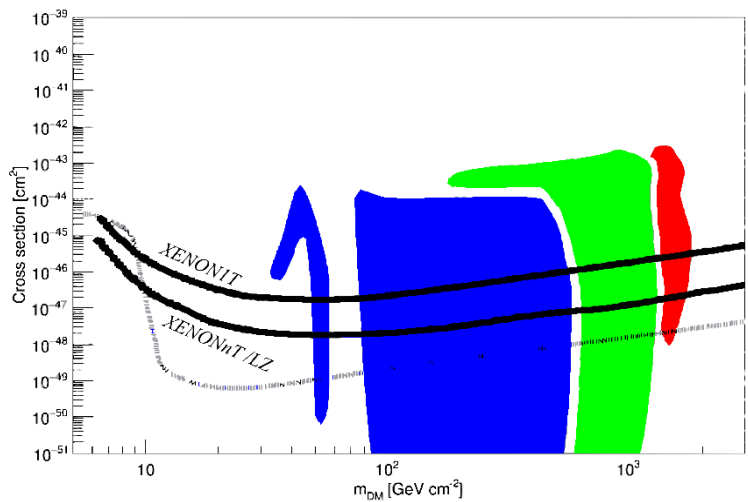

**Figure 4:** The present and future search capabilities on spin-independent WIMP-nucleon scattering (adapted from Chahill-Rawley et al. 2015). The expectation from direct detections (XENON1T/XENON-nT/LZ) overlaid on regions preferred by supersymmetric models. The blue color indicates a regime where the LHC will drive more sensitive constraints, the green represents a region where collider-based searches are complemented by indirect searches by the Cherenkov Telescope Array, and the red depicts a regime most sensitively probed by the Cherenkov Telescope Array. The lower bound on search capabilities imposed by resonant neutrino scattering (light grey) will become the fundamental obstacle for exploring the GeV mass scale window with direct detection. Whereas consistent SUSY models providing WIMPs exist at multi-TeV masses and below the neutrino background they may not constitute the most favoured parameter regions.

Data from Aprile et al. [XENON1T] 2016, Akerib et al. [LZ] 2015

## Where are we heading?

A comparative discussion of the existing experimental approaches can hardly be done without relying on some theoretical preference. If we concentrate on the WIMP, the masses range from a few GeV to 100 TeV. If we further restrict ourselves to Supersymmetric WIMPs, multi-TeV WIMPs are generically disfavoured, but on the other hand WIMPs up to 100 GeV are already significantly constrained by the gamma-ray data. Future data obtained by the Fermi-LAT, direct detection experiments and the LHC will effectively probe the sub-TeV range for WIMP dark matter. The appearance of an unequivocal line signature that will directly reveal the mass of the elusive dark matter particle pronounces line-searches as the least ambiguous among all indirect methods.

Meanwhile, however, line scans increasingly face the inconvenience that data sets are already huge and further enlargements with the current instrumentation are not expected to substantially improve beyond the present sensitivity limits. Only if a significantly better energy resolution in the high-energy gamma-ray experiments will become an experimental reality, things might change. Indeed, there are a number of satellite projects that will provide energy resolutions close to one percent. Outside the WIMP paradigm, though, at similar interaction strength, but predicting somewhat lower masses is Asymmetric Dark Matter (Kaplan et al. 2009). Line searches will also see further application as soon as different energy scales, either towards lower (MeV) or towards higher energies (PeV) will become accessible. Until then, the most promising investigations are consequently those where the number of objects studied can be substantially enlarged (Bechtol et al. [DES] 2015, Drlica-Wagner et al. [DES] 2015, Laevens et al. [PS1] 2015) and translated into more stringent bounds (Drlica-Wagner et al. [DES] 2015, Albert et al. [Fermi-LAT/DES] 2017). Likewise, discoveries of more strongly dark matter-dominated objects, perhaps also located in regions without substantial contributions from the astrophysical backgrounds, presently constitute the most promising way to drive indirect searches beyond the existing constrains. Gaining from the deep cosmological surveys (most notably by the Panoramic Survey Telescope and Rapid Response System, Dark Energy Survey, Large Synoptic Survey Telescope), dark matter subhalos/dwarf spheroidal galaxies appear to be the class of objects that holds the best promise to go beyond even the most stringent current limits. For controlling the systematics, the most promising venue to constrain the extragalactic J-factors is probably related to gravitational lensing investigations (Dalal & Kochanek 2002). As for the Galactic J-factors, current and upcoming Galactic stellar surveys such as GAIA (Perryman et al. 2001) might help to improve the situation.

The privilege to accommodate all indirect search techniques using GeV-scale gamma-rays has over the last decade elevated the Fermi-LAT as the most widely used instrument for indirect searches. The steady accumulation of exposure and gradual improvement of the instrumental response functions as well as the application of elaborate analysis techniques underline the impact of studies utilizing the Fermi-LAT data. At higher energies, the present generation of atmospheric Cherenkov telescopes —most notably H.E.S.S., MAGIC and VERITAS— have contributed decisively to dark matter searches. Follow-up instruments with largely improved sensitivity and/or dynamic range will go beyond these accomplishments. The extended energy range of the High-Altitude Water Cherenkov Observatory (HAWC), the Large High Altitude Air Shower Observatory (LHAASO), and in particular, the Cherenkov Telescope Array (CTA) will allow for searches in a regime presently underexplored or not accessible at all. This appears to be the unique upcoming discovery window for indirect dark matter searches, relating to dark matter candidates on the rather heavy side of the mass scale. These facilities will be able to exceed the present upper bounds of indirect searches using gamma-ray photons. Search opportunities at lower energies are obviously hampered by the void of space instruments since the Compton Gamma-ray Observatory from the era of NASA's large telescopes. Proposed new low-threshold gamma ray telescopes like eASTROGAM or COMPARE might provide prospects for dark matter detection in this underexplored energy regime.

The cosmic ray channel has been most effectively used over the last decade by the Payload for Antimatter Matter Exploration and Light-nuclei Astrophysics (PAMELA) satellite. The Alpha Magnetic Spectrometer (AMS) has by now taken over the frontier of dark matter searches in cosmic rays and will dominate studies in the cosmic ray channel for a substantial period of time. Instruments like the DArk Matter Particle Explorer (DAMPE), the CALorimetric Electron Telescope (CALET), and the High Energy cosmic-Radiation Detection (HERD) hold promise to advance beyond the state-of-the-art in their investigations of signatures in the electron and gamma energy spectra.

We have discussed the potential for indirect detection techniques. If dark matter particle candidates are discovered by the LHC, astrophysical detection will be necessary to connect the produced particle(s) with the cosmic dark matter. Direct probes dark matter on Galactic scales. Indirect detection (including probes in other wavelengths such as optical, radio and microwaves), on the other hand, is the only technique that can probe the particle nature of dark matter on cosmic scales and will continue to do so..

Any attempt to compare the reach of different approaches remains necessarily model-dependent. When the Supersymmetric WIMP is considered the result is usually largely dependent on the implementation of the supersymmetry breaking mechanisms and other choices. Attempts to facilitate the comparison in a more general framework, either using an effective field theory or the so-called simplified models, that reduce the dark matter model to be described by two couplings, a mediator mass and the dark matter mass, are underway but largely lacking for indirect detection. With this caveat in mind we illustrate in Fig.4 the reach of indirect detection (focusing on CTA) as compared to future direct searches, and the parameter space that will be covered by collider-based searches at the LHC (Run 1 at 8 TeV and predictions for Run 2 at 14 TeV in 2023), within the rather generic framework of the 19-dimensional phenomenological Minimal Sypersymmetric Model (pMSSM) e.g. (Chahill-Rawley et al. 2015). It becomes evident that, firstly, there is an ultimate limit that will decisively hamper direct search methods: the background expected from coherent neutrino scattering. Many potential realisations of the pMSSM fall below this boundary, even if admittedly most of these do not really easily qualify for being called 'miracle WIMP' as they will not generically dominate the dark matter and arguably are not favoured by current data. Secondly, for WIMP masses above about a TeV only direct detection and indirect detection have significant discovery potential, and thirdly, among these high mass WIMP models there are those that are within reach only for CTA. Considering time-lines for current and future experiments as well as robustness of predictions, it appears that the unique capabilities for indirect detections are to be found at WIMP masses above a few TeV, that is masses that are not generically preferred by the naturalness requirements on Supersymmetry, but certainly allowed by the WIMP paradigm. From another perspective, however, at masses between about 100 and a thousand GeV there seems to be a non-negligible chance for detection by different search techniques, which would not only provide the necessary confirmation of LHC discoveries but also help to constrain the properties of the dark matter particle. It remains to be seen if indirect search methods will continue to play their role like they have over the last decade.

Given the increasing pressure on the thermally produced WIMP dark matter, especially considering future direct and indirect detection, other particle candidates will have to be considered on equal footing as potential WIMP candidates. Among the foremost of these is probably the 'QCD axion' (Peccei & Quinn 1997, Weinberg 1978, Wilczek 1978) to solve a fine-tuning problem related to CP violation in the QCD sector. Despite the prediction of QCD axions coupling to gamma-rays, the QCD axion, which has a predicted relationship between coupling and mass, is not likely within the reach of indirect detection experiments. Relaxing the model space to axion-like particles (which are a feature of many models beyond the Standard Model), gamma-ray observations are providing constraints which are competitive with the present and planned laboratory experiments (Ajello et al. [Fermi-LAT] 1016, Meyer et al. 2017).

Correspondence should be addressed to O.R. email: olaf.reimer@uibk.ac.at